\documentclass[intlimits,twoside,a4paper]{article}
\usepackage{color} 

\usepackage[cp1251]{inputenc}

\usepackage[eqsecnum]{cmpj3}

\usepackage{bm}

\issue{2018}{21}{3}{33705}
\doinumber{10.5488/CMP.21.33705}

\title[The spin-$\frac{1}{2}$ Heisenberg ferromagnet on the pyrochlore lattice]%
{The spin-$\frac{1}{2}$ Heisenberg ferromagnet on the pyrochlore lattice: A Green's function study%
}
\author[T. Hutak \textsl{et al.}]{T. Hutak\refaddr{label1}, P. M\"{u}ller\refaddr{label2}, J. Richter\refaddr{label2,label3}, T. Krokhmalskii\refaddr{label1}, O. Derzhko\refaddr{label1,label3}}
\addresses{
\addr{label1} Institute for Condensed Matter Physics of the
        National Academy of Sciences of Ukraine,\\
        1 Svientsitskii St., 79011 Lviv, Ukraine
\addr{label2} Institut f\"{u}r theoretische Physik, Otto-von-Guericke-Universit\"{a}t Magdeburg,\\
        P.O. Box 4120, 39016 Magdeburg, Germany
\addr{label3} Max-Planck-Institut f\"{u}r Physik komplexer Systeme, 
        N\"{o}thnitzer Stra\ss e 38, 01187 Dresden, Germany
}

\date{Received June 1, 2018, in final form July 25, 2018}

\begin{document}

\maketitle

\begin{abstract}
We consider the pyrochlore-lattice quantum Heisenberg ferromagnet and discuss the properties of this spin model at arbitrary temperatures.
To this end, we use the Green's function technique within the random-phase (or Tyablikov) approximation 
as well as the linear spin-wave theory and quantum Monte Carlo simulations.
We compare our results to the ones obtained recently by other methods to corroborate our findings.
Finally, we contrast our results with the ones for the simple-cubic-lattice case: 
both lattices are identical at the mean-field level.
We demonstrate that thermal fluctuations are more efficient in the pyrochlore case
(finite-temperature frustration effects).
Our results may be of use for interpreting experimental data for ferromagnetic pyrochlore materials.
\pacs 75.10.-b, 75.10.Jm  
\end{abstract}

The Green's function method in quantum many-body theory emerged in the second half of the twentieth century.
At that time, several outstanding books and reviews such as the ones by 
Kadanoff and Baym \cite{Kadanoff1962}, 
Abrikosov, Gorkov, and Dzyaloshinski \cite{Abrikosov1963},
Zubarev \cite{Zubarev1960}, 
Bonch-Bruevich and  Tyablikov \cite{Bonch-Bruevich1962},
Elk and Gasser \cite{Elk1979}
etc.
appeared.
Professor I.V.~Stasyuk was among the pioneers in Ukraine who began to use these new methods of quantum statistics in solid state researches.
His Green's function papers on ferroelectric and magnetic crystals, molecular excitons or on diagram technique for the Hubbard model 
written in the late sixties--early seventies
are well known in the community \cite{Stasyuk_1a,Stasyuk_1b,Stasyuk_1c,Stasyuk_1d,Stasyuk_1e}.
Moreover, he was lecturing for graduates and post-graduates for many years.
Later on he wrote an excellent textbook on Green's functions \cite{Stasyuk_2}
where one can find many details of calculations as well as various applications in solid state physics.

In what follows we consider one specific quantum spin-lattice system 
--- the spin-$\frac{1}{2}$ Heisenberg pyrochlore ferromagnet ---
using the Green's function formalism.
Clearly, for about sixty years a lot of results were obtained for various quantum spin-lattice systems. 
However, the Green's function technique, even with utilizing a simple random-phase (or Tyablikov) approximation,
is still used in recent studies for exploring complicated spin-lattice models 
(see, e.g., \cite{Vladimirov2016,Mi2016,Sun2018}).
Moreover, while writing this paper we also bear in mind a pedagogical goal 
intending to explain all calculations and compare our findings with the known outcomes of other approaches.
Last but not least, the model we discuss is related to frustrated quantum magnets --- a hot topic during last thirty years. 

It is our pleasure to dedicate this paper to the upcoming eighties birthday of Professor I.V.~Stasyuk,
who has influenced many studies in Lviv, including our ones, over about last fifty years.

\section{Introduction}
\label{sec1}
\setcounter{equation}{0}

One of the standard three-dimensional models for a study of geometrically frustrated spin systems 
is the pyrochlore Heisenberg antiferromagnet \cite{Schollwock2004,Lacroix2011}.
It is known that the nearest-neighbor exchange interactions, because of the lattice geometry 
(a three-dimensional network of corner-sharing tetrahedra),
cannot establish magnetic ordering even at zero temperature $T=0$.
Besides the purely theoretical interest in the spin-system physics of this celebrated model,
there are various real materials 
which can be described with the help of the antiferromagnetic Heisenberg model on the pyrochlore lattice \cite{Gardner2010,Gingras2014}.

Less attention has been paid to the pyrochlore-lattice Heisenberg {\em ferromagnet}.
Clearly, the set of the eigenvalues of the Heisenberg Hamiltonian (energy levels) does not depend on the sign of the exchange interaction.
However, the ordering of these energy levels is different for the antiferromagnetic and ferromagnetic signs of the exchange interaction.
As a result,
the low-energy levels for the pyrochlore-lattice Heisenberg antiferromagnet 
are the high-energy levels for the pyrochlore-lattice Heisenberg ferromagnet
and they may manifest themselves only in finite-temperature thermodynamics,
when due to thermal fluctuations they can come into play.
Several systematic theoretical studies \cite{Mueller2017,Iqbal2018} demonstrated that this is really the case.
For example,
it was found 
that the Curie temperature for the pyrochlore lattice is lower than for the simple-cubic lattice \cite{Garcia-Adeva2014,Mueller2017,Iqbal2018},
although both lattices have the same number of nearest neighbors and thus are identical at the mean-field level.

From the experimental side,
there is a number of compounds, 
the magnetic properties of which can be described using the pyrochlore-lattice quantum Heisenberg ferromagnet 
\cite{Menyuk1966,Wojtowicz1969,Yasui2003,Yaresko2008,Mena2014,Tymoshenko2018}.

In what follows, 
we first introduce the model and present the corresponding linear spin-wave theory (section~\ref{sec2}).
In section~\ref{sec3}, we describe the Green's function approach to the model
and in section~\ref{sec4} we discuss the low- and intermediate-temperature properties of the model making connections to previous studies.
In section~\ref{sec5}, we close the paper with a short summary pointing out some related problems which can be discussed by the same approach. 

\section{The model and the linear spin-wave theory}
\label{sec2}

We begin with a brief description of the model.
The three-dimensional pyrochlore lattice can be viewed as four interpenetrating face-centered-cubic sublattices.
We set the edge length of the cubic cell of each face-centered-cubic sublattice to unity.
Furthermore,
we denote the origins of these face-centered-cubic sublattices as
${\bf{r}}_1=(0,0,0)$,
${\bf{r}}_2=(0,\frac{1}{4},\frac{1}{4})$,
${\bf{r}}_3=(\frac{1}{4},0,\frac{1}{4})$,
and
${\bf{r}}_4=(\frac{1}{4},\frac{1}{4},0)$.
For each face-centered-cubic sublattice, the position of its lattice sites is given by three integers $m_1$, $m_2$, and $m_3$,
i.e., 
by ${\bf{R}}_m=m_1{\bf{e}}_1+m_2{\bf{e}}_2 +m_3{\bf{e}}_3$
with 
${\bf{e}}_1=(0,\frac{1}{2},\frac{1}{2})$,
${\bf{e}}_2=(\frac{1}{2},0,\frac{1}{2})$,
and
${\bf{e}}_3=(\frac{1}{2},\frac{1}{2},0)$,
see \cite{fcc,Mueller2017}.
As a result, 
the lattice sites of the $N$-site pyrochlore lattice are given by
${\bf{R}}_{m;\alpha}={\bf{R}}_m+{\bf{r}}_\alpha$,
where $m=1,\ldots,{\cal{N}}$, ${\cal{N}}=\frac{1}{4}N$ and $\alpha=1,2,3,4$.

We consider the spin-$\frac{1}{2}$ isotropic Heisenberg model.
Its Hamiltonian has the following form: 
\begin{align}
H&=J\sum_{\langle m;\alpha,n;\beta \rangle}{\bf{s}}_{m;\alpha}\cdot {\bf{s}}_{n;\beta}
\nonumber\\
&=
J\sum_{m}
\left( 
{\bf{s}}_{m;1}\cdot {\bf{s}}_{m;2}
+{\bf{s}}_{m;1}\cdot {\bf{s}}_{m;3}
+{\bf{s}}_{m;1}\cdot {\bf{s}}_{m;4}
+{\bf{s}}_{m;2}\cdot {\bf{s}}_{m;3}
+{\bf{s}}_{m;2}\cdot {\bf{s}}_{m;4}
+{\bf{s}}_{m;3}\cdot {\bf{s}}_{m;4}
\right.
\nonumber\\
&\left.
+{\bf{s}}_{m;1}\cdot {\bf{s}}_{{m_1-1,m_2,m_3;2}}
+{\bf{s}}_{m;1}\cdot {\bf{s}}_{{m_1,m_2-1,m_3;3}}
+{\bf{s}}_{m;1}\cdot {\bf{s}}_{{m_1,m_2,m_3-1;4}}
\right.
\nonumber\\
&\left.
+{\bf{s}}_{m_1-1,m_2,m_3;2}\!\cdot\!{\bf{s}}_{{m_1,m_2-1,m_3;3}}
\!+\!{\bf{s}}_{m_1-1,m_2,m_3;2}\!\cdot\!{\bf{s}}_{{m_1,m_2,m_3-1;4}}
\!+\!{\bf{s}}_{m_1,m_2-1,m_3;3}\!\cdot\!{\bf{s}}_{{m_1,m_2,m_3-1;4}}
\right),
\label{201}
\end{align}
where the first sum in equation~(\ref{201}) runs over all nearest neighbors $m;\alpha$ and $n;\beta$,
$J=-\vert J\vert <0$ is the ferromagnetic exchange interaction constant,
and
the sum in the second line in equation~(\ref{201}) runs over all $m=(m_1,m_2,m_3)$, $m_a=1,\ldots,{\cal{L}}_a$ ($a=1,2,3$), ${\cal{L}}_1{\cal{L}}_2{\cal{L}}_3={\cal{N}}$.
The spin operators at different sites commute  and satisfy the following commutation relations at the same site:
\begin{eqnarray}
\label{202}
\left[s^+, s^-\right]=2s^z,
\qquad
\left[s^\pm, s^z\right]=\mp s^\pm,
\end{eqnarray}
where $s^\pm=s^x\pm {\rm{i}}s^y$.
We have assumed $\hbar=1$ for brevity.

We begin with a brief description of the linear spin-wave theory which should be valid in the low-temperature limit.
Using the Holstein-Primakoff transformation
\begin{eqnarray}
\label{203}
s^+=\sqrt{2s}\sqrt{1-\frac{a^\dagger a}{2s}}a\approx \sqrt{2s} a,
\qquad
s^-=\sqrt{2s}a^\dagger \sqrt{1-\frac{a^\dagger a}{2s}}\approx \sqrt{2s} a^\dagger,
\qquad
s^z=s-a^\dagger a,
\end{eqnarray}
we introduce the bosonic operators $a_{m;\alpha}$ (and $a^\dagger_{m;\alpha}$) with $\alpha=1,2,3,4$ and $m=1,\ldots,{\cal{N}}$.
Furthermore, 
we introduce the bosonic operators
\begin{eqnarray}
\label{204}
a_{{\bf{q}};\alpha}=\frac{1}{\sqrt{{\cal{N}}}}
\sum_m{\re}^{-{\rm{i}}{\bf{q}}\cdot{\bf{R}}_m} a_{m;\alpha}\,,
\end{eqnarray}
where ${\bf{q}}\cdot{\bf{R}}_m=q_1m_1+q_2m_2+q_3m_3$
and $q_a=2\piup z_a/{\cal{L}}_a$, $z_a=1,\ldots,{\cal{L}}_a$ ($a=1,2,3$).
[Note that $q_1=\frac{1}{2}(q_y+q_z)$, $q_2=\frac{1}{2}(q_x+q_z)$, and $q_3=\frac{1}{2}(q_x+q_y)$.]
Then, the spin Hamiltonian given in equation~(\ref{201}) can be approximately rewritten in the following form:
\begin{eqnarray}
\label{205}
H=12{\cal{N}}s^2J
-2sJ
\sum_{\bf{q}}
\left(
\begin{array}{cccc}
a^\dagger_{{\bf{q}};1} & a^\dagger_{{\bf{q}};2} & a^\dagger_{{\bf{q}};3} & a^\dagger_{{\bf{q}};4}
\end{array}
\right)
{\bf{F}}
\left(
\begin{array}{c}
a_{{\bf{q}};1} \\
a_{{\bf{q}};2} \\
a_{{\bf{q}};3} \\
a_{{\bf{q}};4}
\end{array}
\right),
\end{eqnarray}
where
\begin{eqnarray}
\label{206}
&{\bf{F}}
=
\left(
\begin{array}{cccc}
3 & -\phi_1 & -\phi_2 & -\phi_3 \\
-\phi^*_1 & 3 & -\phi_{12}^* & -\phi_{13}^* \\
-\phi^*_2 & -\phi_{12} & 3 & -\phi_{23}^* \\
-\phi^*_3 & -\phi_{13} & -\phi_{23} & 3
\end{array}
\right),&
\nonumber\\
&\phi_a={\re}^{-\frac{{\rm{i}}q_a}{2}}\cos\frac{q_a}{2}\,,
\qquad
\phi_{ab}={\re}^{-\frac{{\rm{i}}(q_a-q_b)}{2}}\cos\frac{q_a-q_b}{2}.&
\end{eqnarray}
After diagonalizing the bilinear Bose form in equation~(\ref{205}) we arrive at
\begin{eqnarray}
\label{207}
H=12{\cal{N}}s^2J
-2sJ\sum_{\gamma=1}^4\sum_{\bf{q}}
\omega_{\gamma;{\bf{q}}}\xi_{{\bf{q}};\gamma}^\dagger \xi_{{\bf{q}};\gamma}\,.
\end{eqnarray}
The energies of the linear spin waves (magnons) 
$-2sJ\omega_{\gamma;{\bf{q}}}=\vert J\vert \omega_{\gamma;{\bf{q}}}$
are determined by the eigenvalues of the matrix ${\bf{F}}$
\begin{eqnarray}
\label{208}
\omega_{1;{\bf{q}}}=\omega_{2;{\bf{q}}}=4,
\qquad
\omega_{3;{\bf{q}}}=2+D_{{\bf{q}}}\,,
\qquad
\omega_{4;{\bf{q}}}=2-D_{{\bf{q}}}\,,
\nonumber\\
D^2_{{\bf{q}}}
=1+\cos\frac{q_x}{2}\cos\frac{q_y}{2}+\cos\frac{q_x}{2}\cos\frac{q_z}{2}+\cos\frac{q_y}{2}\cos\frac{q_z}{2}.
\end{eqnarray}
The calculated spin-wave dispersion coincides with the result reported in \cite{Mena2014}
[see figure~1~(c) of that paper].

Now, from equation~(\ref{207}) we can obtain the internal (intrinsic) energy per site
\begin{eqnarray}
\label{209}
\frac{\langle H\rangle}{N}
=3s^2J-\frac{sJ}{2{\cal{N}}}\sum_{\gamma=1}^4\sum_{\bf{q}}
\omega_{\gamma;{\bf{q}}}n_{{\bf{q}};\gamma}
\end{eqnarray}
and the magnetization per site
\begin{eqnarray}
\label{210}
\langle s^z\rangle=\frac{1}{N}\sum_{m;\alpha}\langle s^z_{m;\alpha}\rangle
=s-\frac{1}{4{\cal{N}}}\sum_{\gamma=1}^4\sum_{\bf{q}}
n_{{\bf{q}};\gamma}\,.
\end{eqnarray}
Here, 
$n_{{\bf{q}};\gamma}=1/({\re}^{\vert J\vert\omega_{\gamma;{\bf{q}}}/T}-1)$ 
denotes the Bose-Einstein distribution function
(we set $k_{\rm{B}}=1$ for brevity)
and $s=\frac{1}{2}$.
Below we use the linear spin-wave theory predictions (\ref{209}) and (\ref{210}) for comparison with the Green's function approach results.

\section{Green's function method. Random-phase approximation}
\label{sec3}
\setcounter{equation}{0}

Now, we turn to the Green's function approach 
\cite{Zubarev1960,Bonch-Bruevich1962,Elk1979,Stasyuk_2,Tyablikov1967,Gasser2001,Froebrich2006,Majlis2007}.
We introduce the following operators
\begin{eqnarray}
\label{301}
s_{{\bf{q}};\alpha}^{\pm}
=
\frac{1}{\sqrt{{\cal{N}}}}\sum_{m}{\re}^{\mp{\rm{i}}{\bf{q}}\cdot{\bf{R}}_m}s^{\pm}_{m;\alpha}
\end{eqnarray}
and the (retarded) Green's function
\begin{eqnarray}
\label{302}
G_{\alpha\beta}(t)
\equiv
\langle\langle s^+_{{\bf{q}};\alpha}\vert s^-_{{\bf{q}};\,\beta}\rangle\rangle_{t}
=
-{\rm{i}}\Theta(t)\langle [ s^+_{{\bf{q}};\alpha}(t),s^-_{{\bf{q}};\,\beta}]\rangle,
\qquad
G_{\alpha\beta}(\omega)
=
\int_{-\infty}^{\infty}{\rm{d}}t {\rm {e}}^{{\rm{i}}\omega t}G_{\alpha\beta}(t),
\end{eqnarray}
see, e.g., \cite{Tyablikov1967}.
The first-order equation of motion reads:
\begin{eqnarray}
\label{303}
\omega
\langle\langle s^+_{{\bf{q}};\alpha}\vert s^-_{{\bf{q}};\,\beta}\rangle\rangle_{\omega}
=
\frac{2}{{\cal{N}}}\sum_m\langle s_{m;\alpha}^z\rangle \delta_{\alpha\beta}
+
{\rm{i}}\langle\langle \dot{s}^+_{{\bf{q}};\alpha}\vert s^-_{{\bf{q}};\,\beta}\rangle\rangle_{\omega}\,,
\qquad
{\rm{i}}\dot{s}^+_{{\bf{q}};\alpha}
=
\frac{1}{\sqrt{{\cal{N}}}}\sum_m {\re}^{-{\rm{i}}{\bf{q}}\cdot {\bf{R}}_m}
\left[s^+_{m;\alpha},H\right].
\end{eqnarray}
Calculating the commutator in the right-hand side of the second equation in equation~(\ref{303}) 
(see appendix)
and introducing the random-phase (or Tyablikov) approximation
$s^z_{\text A}s^\pm_{\text B}\to \langle s^z\rangle s^\pm_{\text B}$ \cite{Tyablikov1967}
(see appendix)
we obtain the closed-form equation for the Green's function
\begin{eqnarray}
\label{304}
\sum_\gamma\left(\omega\delta_{\alpha\gamma}
+
2J\langle s^z\rangle F_{\alpha\gamma}\right)G_{\gamma\beta}(\omega)
=
2\langle s^z\rangle\delta_{\alpha\beta}\,,
\end{eqnarray}
where 
$-2J\langle s^z\rangle{\bf{F}}$ is the frequency matrix, 
${\bf{G}}(\omega)$ is the matrix of Green's functions,
and 
${\bf{M}}=2\langle s^z\rangle {\bf{1}}$ is the moment matrix,
see also equation~(\ref{a04}).
The matrix ${\bf {F}}$ appeared already in the linear spin-wave calculations,
see equation~(\ref{206}).

We can find the eigenvalues and the eigenvectors of the matrix ${\bf{F}}$,
which satisfy 
$\sum_\beta F_{\alpha\beta}\langle\beta \vert\gamma{\bf{q}}\rangle
=\omega_{\gamma;{\bf{q}}}\langle \alpha \vert\gamma{\bf{q}}\rangle$.
The eigenvalues $\omega_{\gamma;{\bf{q}}}$, $\gamma=1,2,3,4$ are given in equation~(\ref{208}).
The (orthonormal) eigenvectors $\langle \alpha \vert\gamma{\bf{q}}\rangle$, $\gamma=1,2,3,4$ are too lengthy to be presented here.

Finally, 
using the identity
${\bf{1}}=\sum_\gamma \vert \gamma{\bf{q}}\rangle\langle \gamma{\bf{q}}\vert$,
we can resolve equation~(\ref{304}) with respect to ${\bf{G}}$,
${\bf{G}}=2\langle s^z\rangle(\omega {\bf{1}}+2J\langle s^z\rangle{\bf{F}})^{-1}\sum_\gamma \vert \gamma{\bf{q}}\rangle\langle \gamma{\bf{q}}\vert
=\sum_\gamma[2\langle s^z\rangle/(\omega+2J\langle s^z\rangle\omega_{\gamma;{\bf{q}}})]\vert \gamma{\bf{q}}\rangle\langle \gamma{\bf{q}}\vert$,
to get the desired result
\begin{eqnarray}
\label{305}
G_{\alpha\beta}(\omega)
=
2 \langle s^z\rangle
\sum_{\gamma=1}^4\frac{\langle\alpha\vert \gamma{\bf{q}}\rangle\langle \gamma{\bf{q}}\vert\beta\rangle}
{\omega+2J\langle s^z\rangle\omega_{\gamma;{\bf{q}}}}.
\end{eqnarray}
Here, $\langle \gamma{\bf{q}}\vert\beta\rangle$ is the $\beta$-th component of the eigenvector $\langle \gamma{\bf{q}}\vert$.
Clearly, the magnetization $\langle s^z\rangle$ which enters equation~(\ref{305}) should be determined self-consistently,
see equation~(\ref{307}) below.

As can be seen from equation~(\ref{305}),
the excitation energies are given by 
$-2J\langle s^z\rangle \omega_{\gamma;{\bf{q}}}=2\vert J\vert\langle s^z\rangle \omega_{\gamma;{\bf{q}}}$.
In the limit $T\to 0$,
when $\langle s^z\rangle\to \frac{1}{2}$,
they coincide with the magnon energies considered in section~\ref{sec2}
and, therefore, we call these excitations magnons.
Moreover, they coincide with the results of \cite{Mueller2017} in this limit
(see the upper panel in figure~2 of that paper).
For finite temperatures $T>0$ the dependences on ${\bf{q}}$ remain unchanged,
and the energies simply decrease due to the factor $\langle s^z\rangle<\frac{1}{2}$.
More sophisticated effects of the temperature on the excitation energies 
were discussed 
within the rotation-invariant (or Kondo-Yamaji \cite{Kondo1972}) Green's function approach in \cite{Mueller2017}.\footnote{The rotation-invariant Green's function approach 
considers the equation of motion up to the second order,
which obviously involves Green's functions of higher order than the initial ones.
Several products of three-spin operators are then simplified by a decoupling like
$s^+_{\text A} s^-_{\text B} s^z_{\text C} \to \alpha_\text{AB} \langle s^+_{\text A} s^-_{\text B} \rangle s^z_{\text C}$ etc.,
where $\alpha_\text{AB}$ is a vertex parameter which is introduced to improve the approximation made by decoupling.
After all,
the correlation functions, the vertex parameter, and the condensation term which is related to magnetic long-range order
are determined  self-consistently
[within the Tyablikov approximation,
one faces only one resulting equation for $\langle s^z \rangle$, see equation~(\ref{307})].
For further details see \cite{Mueller2017}.}
Since $D_{\bf{q}}\to 2-\frac{1}{16}\vert{\bf{q}}\vert^2$ as ${\bf{q}}\to 0$,
$\omega_{4;{\bf{q}}}$ is the acoustic branch of the spectrum,
i.e., $\omega_{4;{\bf{q}}}\to\rho\vert{\bf{q}}\vert^2$, 
$\rho=\frac{1}{8}\vert J\vert\langle s^z\rangle$ as ${\bf{q}}\to 0$.
The spin stiffness $\rho$ is proportional to $\langle s^z\rangle$;
it shows the same temperature behavior as $\langle s^z\rangle$ vanishing as $T\to T_\text{c}$.

Knowing the Green's functions, 
we calculate the equal-time correlation functions via the relation
\begin{eqnarray}
\label{306}
\langle s^-_{{\bf{q}};\,\beta} s^+_{{\bf{q}};\alpha}\rangle
=
\frac{{\rm{i}}}{2\piup}\lim_{\epsilon\to +0}\int_{-\infty}^{\infty} {\rm{d}}\omega
\frac{G_{\alpha\beta}(\omega+{\rm{i}}\epsilon)-G_{\alpha\beta}(\omega-{\rm{i}}\epsilon)}{{\re}^{\frac{\omega}{T}}-1}
=
2 \langle s^z\rangle 
\sum_{\gamma=1}^4
\frac{\langle\alpha\vert \gamma{\bf{q}}\rangle\langle \gamma{\bf{q}}\vert\beta\rangle}
{{\re}^{-\frac{2J\langle s^z\rangle\omega_{\gamma;{\bf{q}}}}{T}}-1}.
\end{eqnarray}
These correlation functions yield important thermodynamic characteristics.
In particular,
using the identity $s^z=\frac{1}{2}-s^-s^+$ which leads to
$\langle s^z\rangle=\frac{1}{2}-\frac{1}{N}\sum_{m;\alpha}\langle s^-_{m;\alpha} s^+_{m;\alpha}\rangle
=\frac{1}{2}-\frac{1}{N}\sum_\alpha\sum_{\bf{q}}\langle s^-_{{\bf{q}};\alpha} s^+_{{\bf{q}};\alpha}\rangle$, 
we arrive at the following formula for $\langle s^z\rangle$:
\begin{eqnarray}
\label{307}
\langle s^z\rangle
=
\frac{1}{2}
-\frac{\langle s^z\rangle}{2{\cal{N}}}\sum_{\gamma=1}^4\sum_{\bf{q}}
\frac{1}{{\re}^{-\frac{2J\langle s^z\rangle\omega_{\gamma;{\bf{q}}}}{T}}-1}.
\end{eqnarray}
In fact, 
this is the equation for determining self-consistently the magnetization $\langle s^z\rangle$.

We can obtain the internal energy of the spin model at hand by averaging the Hamiltonian (\ref{201}).
Since we have calculated only the Green's function $\langle \langle s^+\vert s^-\rangle\rangle$ 
rather than $\langle \langle s^z\vert s^z\rangle\rangle$
we should eliminate the correlations $\langle s^z s^z\rangle$ from the consideration
in order to treat all correlations in $\langle H\rangle$ on equal footing,
i.e., within the same (Tyablikov) approximation \cite{Froebrich2006,Rutonjski2011}. 
After some manipulations
(see appendix),
we arrive at the following expression for the internal energy within the Tyablikov approximation:
\begin{align}
\frac{\langle H\rangle}{N}
&=\frac{3}{2}J\langle s^z\rangle
-\frac{J\langle s^z\rangle^2}{2{\cal{N}}}\sum_{\gamma=1}^4\sum_{\bf{q}}
\frac{\omega_{\gamma;{\bf{q}}}}{{\re}^{-\frac{2J\langle s^z\rangle\omega_{\gamma;{\bf{q}}}}{T}}-1}
+\frac{J\langle s^z\rangle}{4{\cal{N}}}
\sum_{\gamma=1}^4
\sum_{\substack{\alpha,\beta=1\\(\alpha\ne\beta)}}^4
\sum_{\bf{q}}
\varphi_{\alpha\beta}
\frac{\langle\alpha\vert\gamma{\bf{q}}\rangle \langle\gamma{\bf{q}}\vert\beta\rangle}
{{\re}^{-\frac{2J\langle s^z\rangle\omega_{\gamma;{\bf{q}}}}{T}}-1}\,,
\nonumber\\
\varphi_{12}&=\varphi_{21}^*=\phi_1\,, \qquad
\varphi_{13}=\varphi_{31}^*=\phi_2\,, \qquad
\varphi_{14}=\varphi_{41}^*=\phi_3\,,
\nonumber\\
\varphi_{23}&=\varphi_{32}^*=\phi_{12}^*\,, \qquad
\varphi_{24}=\varphi_{42}^*=\phi_{13}^*\,, \qquad
\varphi_{34}=\varphi_{43}^*=\phi_{23}^*\,,
\label{308}
\end{align}
where $\phi_a$ and $\phi_{ab}$ are defined in equation~\eqref{206}.
Differentiation of the internal energy with respect to $T$ gives the specific heat.

The Green's function (\ref{305}) gives the dynamic susceptibility,
$\chi^{+-}_{\bf{q}}(\omega)\equiv\frac{1}{4}\sum_{\alpha,\beta=1}^4\chi^{+-}_{{\bf{q}};\alpha\beta}(\omega)$,
$\chi^{+-}_{{\bf{q}};\alpha\beta}(\omega)=-G_{\alpha\beta}(\omega)$,
which is related to the dynamic structure factor
\begin{eqnarray}
\label{309}
S^{+-}_{\bf{q}}(\omega)
=
\frac{2}{1-{\re}^{-\frac{\omega}{T}}}\Im \chi^{+-}_{\bf{q}}(\omega)
=\frac{\piup\langle s^z\rangle}{1-{\re}^{-\frac{\omega}{T}}}
\sum_{\alpha,\beta,\gamma=1}^4\langle\alpha\vert{\bf{q}}\gamma\rangle\langle {\bf{q}}\gamma\vert\beta\rangle
\delta\left(\omega-2\vert J\vert \langle s^z\rangle \omega_{\gamma;{\bf{q}}}\right).
\end{eqnarray}

Furthermore, 
the Green's function given in equation~(\ref{305}) is proportional to $\langle s^z\rangle$ 
and at zero field it disappears in the high-temperature paramagnetic phase above the critical (Curie) temperature $T_{\text c}$.
However, the (initial) susceptibility $\chi$, 
defined as $\langle s^z\rangle=\chi h$, where $h$ is an infinitesimally small applied magnetic field can be determined as follows.
For the spin model with the Hamiltonian $H-h\sum_{m;\alpha}s^z_{m;\alpha}$,
the Green's function $G_{\alpha\beta}(\omega)$ within the random-phase approximation is again given by equation~(\ref{305}),
although
with $2J\langle s^z\rangle\omega_{\gamma;{\bf{q}}}-h$ instead of $2J\langle s^z\rangle\omega_{\gamma;{\bf{q}}}$ in the denominator.
Therefore, the equation for $\langle s^z\rangle$ now reads
\begin{eqnarray}
\label{310}
\langle s^z\rangle
=
\frac{1}{2}
-\frac{\langle s^z\rangle}{2{\cal{N}}}\sum_{\gamma=1}^4\sum_{\bf{q}}
\frac{1}{{\re}^{-\frac{2J\langle s^z\rangle\omega_{\gamma;{\bf{q}}}-h}{T}}-1}\,,
\end{eqnarray}
cf. equation~(\ref{307}).
Following, e.g., \cite{Tyablikov1967}, for $T>T_{\text c}$ we set $\langle s^z\rangle=0$ in the left-hand side of equation~(\ref{310})
and
then expand the exponent and substitute $\langle s^z\rangle=\chi h$
to arrive at the equation for $1/\chi$ above the critical temperature $T_{\text c}$:
\begin{eqnarray}
\label{311}
1
=
\frac{1}{{\cal{N}}}\sum_{\gamma=1}^4\sum_{\bf{q}}
\frac{T}{-2J\omega_{\gamma;{\bf{q}}}+\frac{1}{\chi}}.
\end{eqnarray}

All calculations described in this section 
are, in principle, well known \cite{Zubarev1960,Bonch-Bruevich1962,Elk1979,Stasyuk_2,Tyablikov1967,Gasser2001,Froebrich2006,Majlis2007}.
However, to our best knowledge, 
we are not aware of any random-phase Green's function paper for the pyrochlore-lattice quantum Heisenberg ferromagnet.

\section{Low- and intermediate-temperature results}
\label{sec4}
\setcounter{equation}{0}

\begin{figure}[!b]
\centering
\includegraphics[clip=on,width=0.6\textwidth,angle=0]{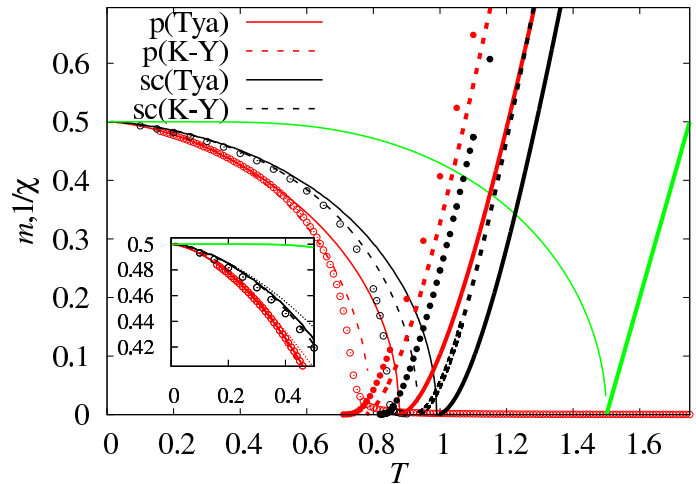}
\caption{(Colour online) Temperature dependence of magnetization $m=\langle s^z\rangle$ (thin) and inverse susceptibility $1/\chi$ (thick) 
for the spin-$\frac{1}{2}$ pyrochlore-lattice (red) Heisenberg ferromagnet ($J=-1$)
obtained within Tyablikov approximation (solid) and rotation-invariant Green's function method (dashed).
For comparison, we also show the results for the simple-cubic lattice (black).
Note that the mean-field data (solid green) are identical for both lattices.
Quantum Monte Carlo simulation data for $N=16\cdot 32^3$ sites (pyrochlore, red) and $N=80^3$ sites (simple-cubic, black)
are shown by open circles ($m$) and filled circles ($1/\chi$).
Inset: Magnetization at low temperatures. 
We also show here the linear spin-wave theory predictions (very thin dotted).}
\label{fig01}
\end{figure}

\begin{figure}[!t]
\centering
\includegraphics[clip=on,width=0.6\textwidth,angle=0]{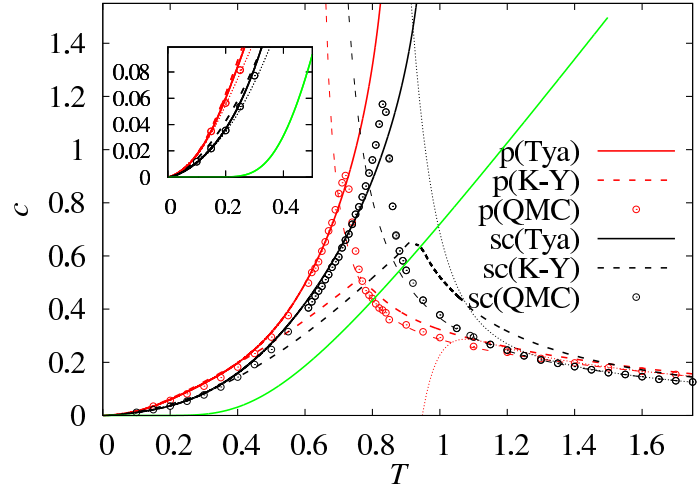}
\caption{(Colour online) Temperature dependence of specific heat per site $c$
for the spin-$\frac{1}{2}$ pyrochlore-lattice (red) Heisenberg ferromagnet ($J=-1$)
obtained within Tyablikov approximation (solid) and rotation-invariant Green's function method (dashed).
We compare the results for the pyrochlore lattice (red) and the simple-cubic lattice (black)
which are identical at the mean-field level (solid green).
We provide in addition quantum Monte Carlo data 
obtained for $N=16\cdot 16^3$ (pyrochlore, red empty circles) and $N=40^3$ (simple-cubic, black empty circles) sites.
Very thin dotted (double-dashed) lines represent the 13th order high-temperature expansion 
raw (extrapolated by the $[6,7]$ Pad\'{e} approximant) results.
Inset: Specific heat at low temperatures.
We also show here the linear spin-wave theory predictions (very thin dotted). 
It is in order to mention here, that for the $c(T)$ curve of the simple-cubic lattice presented in figure~8 of \cite{Mueller2017}, 
an incorrect data set was used.}
\label{fig02}
\end{figure}

In the low-temperature limit, 
the excitations described by $G_{\alpha\beta}(\omega)$ (\ref{302}), (\ref{305}) are magnons also emerging in the linear spin-wave theory.
Consequently,
in the low-temperature limit, the Green's function results are expected to coincide with the ones of the linear spin-wave theory,
see the low-temperature region in figures~\ref{fig01} and \ref{fig02}.

Let us compute the critical (Curie) temperature $T_{\text c}$.
At the critical temperature, $\langle s^z\rangle$ vanishes;
we set $\langle s^z\rangle=0$ in the left-hand side of equation~(\ref{307}) and expand the exponent in the right-hand side of equation~(\ref{307}) 
to obtain the critical temperature:
\begin{eqnarray}
\label{401}
\frac{T_{\text c}}{\vert J\vert}
=\frac{2}{\frac{1}{{\cal{N}}}\sum_{\alpha=1}^4\sum_{{\bf{q}}}\frac{1}{\omega_{\alpha;{\bf{q}}}}}
\approx 0.872.
\end{eqnarray}
Here,
in the thermodynamic limit
$\frac{1}{{\cal{N}}}\sum_{\bf{q}}(\ldots)\to\frac{1}{\piup^3}\int_0^\piup{\rm{d}}q_1\int_0^\piup{\rm{d}}q_2\int_0^\piup{\rm{d}}q_3(\ldots)$.
The same equation follows from equation~(\ref{311}) after setting $1/\chi=0$.
This result should be compared to the corresponding ones obtained by other means given in the second column of table~\ref{tab01}.
The important feature is visible: 
the critical temperature calculated within the random-phase approximation Green's function method 
is the highest one in comparison with the results of other approaches
(excluding mean-field)
indicating some underestimate of the role of thermal fluctuations.
Comparing the results for the pyrochlore and simple-cubic cases in the same approximation,
we again observe a manifestation of finite-temperature frustration effects \cite{Schmalfuss2005,Mueller2017}:
thermal fluctuations destroy the magnetic order more effectively for the pyrochlore-lattice geometry,
and the critical temperature for the pyrochlore ferromagnet is about 15\% smaller than for the simple-cubic ferromagnet.
Of course,
the mean-field result for both lattices is the same: $T_{\text c}=\frac{3}{2}\vert J\vert$.

\begin{table}[!t]
\centering
\caption{Critical temperature $T_{\text c}/\vert J\vert$ 
for the spin-$\frac{1}{2}$ pyrochlore and simple-cubic Heisenberg ferromagnets obtained by 
quantum Monte Carlo (QMC), 
high-temperature expansions (HTE),
pseudofermion functional renormalization group method (PFFRG),
rotation-invariant Green's function method (RGM),
and random-phase approximation Green's function method (RPA).
Within mean-field approximation (MFA, the last row) $T_{\text c}$ for both lattices is identical.}
\label{tab01}
\vspace{2ex}
\renewcommand{\arraystretch}{1.25}
\begin{tabular}{|c|c|c||c|}
\hline\hline
      & pyrochlore lattice                  & simple-cubic lattice                   & $\dfrac{T_{\text c}^{\rm{pyro}\strut}}{T_{\text c}^{\rm{sc}}}$ \\[3mm]
\hline\hline
QMC   & 0.718 \cite{Mueller2017}            & 0.839(1) \cite{Wessel2010}             & 86\% \\
\hline
HTE   & 0.724\,\ldots0.754 \cite{Lohmann2014} & 0.827 \cite{Lohmann2014}               & 88\% \\
\hline
PFFRG & 0.77(4) \cite{Iqbal2018}            & 0.90(4) \cite{Iqbal2018}               & 86\% \\
\hline
RGM   & 0.778 \cite{Mueller2017}            & 0.926 \cite{Junger2009,Menchyshyn2014} & 84\% \\
\hline
RPA   & 0.872 (this paper)                  & 0.989 \cite{Tyablikov1967}             & 88\% \\
\hline
\hline
MFA   & $\frac{3}{2}$                       & $\frac{3}{2}$                          & 100\% \\
\hline
\end{tabular}
\end{table}

We can solve equation~(\ref{307}) with respect to $\langle s^z\rangle$ for all temperatures below $T_{\text c}$ (\ref{401}).
This temperature dependence of the magnetization is shown in figure~\ref{fig01}
along with the results of the rotation-invariant Green's function method \cite{Mueller2017} 
and the quantum Monte Carlo simulations using the ALPS package \cite{ALPS1,ALPS2} (see also \cite{Menchyshyn2014}).
We also report in figure~\ref{fig01} the corresponding results for the simple-cubic case for comparison.
Moreover, in figure~\ref{fig01} we show additionally the temperature dependence of $1/\chi$ in the paramagnetic phase 
as it follows from equation~\eqref{311} (thick solid),
from the rotation-invariant Green's function method \cite{Mueller2017} (thick dashed), 
and from the quantum Monte Carlo simulations (filled circles)
for the pyrochlore (red) and simple-cubic (black) lattices.
Within the mean-field approximation $1/\chi=4(T_{\text c}-T)$, $T_{\text c}=\frac{3}{2}\vert J\vert$ for both lattices,
see thick solid green curve.
The reported results for the temperature dependences of magnetization and susceptibility agree
well with each other
and illustrate the finite-temperature frustration effects as well as the quality of various approximations.

The specific heat within the Tyablikov approximation is evaluated by finding the derivative of equation~(\ref{308}) with respect to temperature $T$.
In figure~\ref{fig02} we compare the temperature dependence of the specific heat obtained by different methods
for the pyrochlore lattice (red) and simple-cubic lattice (black).
The mean-field result for both lattices is the same (green):
the critical temperature $T_{\text c}$ is overestimated and $c(T_{\text c})$ is finite.
The Tyablikov approximation (solid red and solid black) yields already different values for $T_{\text c}$, 
although the specific heat diverges at the Curie temperature.
This contradicts the current critical exponent estimates of the three-dimensional Heisenberg model  
obtained by field theory \cite{Guillou1980}, high-temperature series analyses, and Monte Carlo \cite{Holm1993a,Holm1993b,Souza2000} methods
which predict that the specific heat is not singular for this model.
According to the rotation-invariant Green's function method, $c(T_{\text c})$ is finite (dashed red and dashed black).
We also report quantum Monte Carlo data obtained using the ALPS package \cite{ALPS1,ALPS2} (see also \cite{Menchyshyn2014});
red (black) empty circles refer to the pyrochlore (simple-cubic) lattice of $N=16\cdot 16^3$ ($N=40^3$) sites.
Moreover,
we show the 13th order high-temperature expansion results, 
raw data (very thin dotted) and $[6,7]$ Pad\'{e} approximants (very thin double-dashed),
obtained using the HTE program freely available at \url{http://www.uni-magdeburg.de/jschulen/HTE},
see \cite{Lohmann2011,Lohmann2014}.
In summary,
the presented data show that the Tyablikov approximation is capable of reproducing reasonably well the dependence $c(T)$ in the ferromagnetic phase,
except the region close to $T_{\text c}$.
In the paramagnetic phase, 
the rotation-invariant Green's function method and, of course, high-temperature expansion method give reasonably good results.

\begin{figure}[!t]
\centering
\includegraphics[clip=on,width=0.55\textwidth,angle=0]{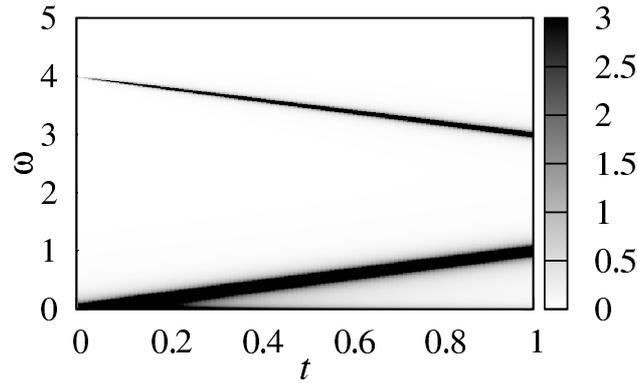}
\caption{Dynamic structure factor $S^{+-}_{{\bf{q}}}(\omega)$
for the spin-$\frac{1}{2}$ pyrochlore-lattice Heisenberg ferromagnet ($J=-1$)
as a function of the reduced momentum $t=2-D_{\bf{q}}$ with ${\bf{q}}=(q,q,q)$ at $T=0.0425$
obtained within the Tyablikov approximation.
$\delta$-function is smeared by introducing the Lorentzian function with finite $\epsilon=0.01$.}
\label{fig03}
\end{figure}

Finally,
we turn to the dynamic structure factor $S^{+-}_{{\bf{q}}}(\omega)$.
Within the Tyablikov approximation, it is given by equation~(\ref{309}).
For further calculations we replace in equation~(\ref{309})
$\delta(x)$ by the Lorentzian function $\epsilon/[\piup(x^2+\epsilon^2)]$ with $\epsilon=0.01$.
Inspired by the experimental paper on the spin-$\frac{1}{2}$ pyrochlore ferromagnet Lu$_2$V$_2$O$_7$ \cite{Mena2014},
where neutron inelastic scattering data were reported (see figure~2 of that paper),
we calculate $S^{+-}_{{\bf{q}}}(\omega)$ (\ref{309}) at $T=0.0425\vert J\vert$ along  the path ${\bf{q}}=(q,q,q)$ 
and present the result as a function of the reduced momentum $t=2-D_{\bf{q}}$,
see figure~\ref{fig03}.
(Experimentalists collected data from more points in the ${\bf{q}}$-space, not only along the path ${\bf{q}}=(q,q,q)$, see \cite{Mena2014}.)
Overall, figure~\ref{fig03} resembles the experimental data reported in \cite{Mena2014};
the dynamic structure factor is concentrated along the lines 
$\omega_{4;{\bf{q}}}=4-t$, 
$\omega_{3;{\bf{q}}}=t$, 
however,
because of the chosen path ${\bf{q}}=(q,q,q)$,
not along the line
$\omega_{1;{\bf{q}}}=\omega_{2;{\bf{q}}}=4$.
While comparing with experiments (and with the rotation-invariant Green's function result \cite{Mueller2017}),
one should remember that $S^{+-}_{{\bf{q}}}(\omega)$ presents only the transverse part, 
while the longitudinal part $S^{zz}_{{\bf{q}}}(\omega)$ is not taken into consideration.
We note in passing
that comparing theoretical predictions with experimental data one can obtain the exchange interaction constants,
and the exploited here random-phase approximation Green's function method can be used for this purpose.

\section{Summary}
\label{sec5}
\setcounter{equation}{0}

In the present paper
we have discussed the thermodynamic properties of the pyrochlore-lattice quantum Heisenberg ferromagnet.
Our main results 
(Curie temperature, 
temperature dependences of the magnetization and the specific heat, 
dynamic structure factor)
have been obtained within the random-phase approximation.
We compare them to the linear spin-wave theory at low temperatures 
as well as to the rotation-invariant Green's function method \cite{Mueller2017},
quantum Monte Carlo simulation, and high-temperature expansion \cite{Lohmann2014,Mueller2017} results.
Although the random-phase approximation goes beyond the mean-field treatment, 
thus leading to different results for the pyrochlore-lattice and the simple-cubic-lattice case, 
it takes into account not all fluctuations, and ferromagnetic ordering is slightly overestimated
in comparison with the second-order rotation-invariant Green's function method results.
Nevertheless, 
the random-phase approximation Green's function method together with the high-temperature expansion 
provide a simple route to describe quantum Heisenberg ferromagnets for all temperatures.
Of course, the quantum Monte Carlo simulations or the rotation-invariant Green's function method are also applicable,
though these are more complicated techniques.

The Green's function method which uses the Tyablikov approximation can be applied to some other spin models related to the pyrochlore lattice.
For example, 
one can in a similar way consider the ferromagnet 
in the presence of the breathing anisotropy or the second-nearest neighbor interactions.
However,
the most interesting problem within this context  
is the study of the properties of the pyrochlore-lattice quantum Heisenberg antiferromagnet.
In zero magnetic field,
the partition functions of the Heisenberg ferro- and antiferromagnet are connected by the obvious relation: 
$Z_{\rm{FM}}(-T)=Z_{\rm{AFM}}(T)$.
This implies that the specific heat of the antiferromagnet at temperature $T$ is given by the specific heat of the ferromagnet taken at temperature $-T$.
However, in the thermodynamic limit, because of spontaneous symmetry breaking, this relation is not valid.
Clearly,
the rotation-invariant Green's function approach, 
which implies $\langle s^z\rangle=0$, 
seems to be an appropriate approximation for examining that case.
The work in this direction is in progress \cite{Mueller2018}.

\section*{Acknowledgements}%

We thank O.~Menchyshyn for collaboration on the early stage of this project.
O.D. acknowledges the kind hospitality of the MPIPKS, Dresden in April-June of 2018.

\appendix
\section{First-order equation of motion, Tyablikov approximation, and some other calculations}

To obtain the first-order equation of motion we have to calculate the commutator in the second equation in equation~(\ref{303}). 
We begin with the case $\alpha=1$ in equation~(\ref{303}).
Then,
\begin{align}
\frac{1}{J}
\left[s_{m;1}^+,H\right]
&=
\sum_n
\Big[
s_{m;1}^+,
\frac{1}{2}s_{n;1}^-\left(s_{n;2}^+ + s_{n;3}^+ + s_{n;4}^+\right)
+s_{n;1}^z\left(s_{n;2}^z + s_{n;3}^z + s_{n;4}^z\right)
\nonumber\\
&+\frac{1}{2}s_{n;1}^-\left(s_{n_1-1,n_2,n_3;2}^+ + s_{n_1,n_2-1,n_3;3}^+ + s_{n_1,n_2,n_3-1;4}^+\right)
\nonumber\\
&+s_{n;1}^z\left(s_{n_1-1,n_2,n_3;2}^z + s_{n_1,n_2-1,n_3;3}^z + s_{n_1,n_2,n_3-1;4}^z\right)
\Big]
\nonumber\\
&=
s_{m;1}^z\left(s_{m;2}^+ + s_{m;3}^+ + s_{m;4}^+\right)
-s_{m;1}^+\left(s_{m;2}^z + s_{m;3}^z + s_{m;4}^z\right)
\nonumber\\
&+s_{m;1}^z\left(s_{m_1-1,m_2,m_3;2}^+ + s_{m_1,m_2-1,m_3;3}^+ + s_{m_1,m_2,m_3-1;4}^+\right)
\nonumber\\
&-s_{m;1}^+\left(s_{m_1-1,m_2,m_3;2}^z + s_{m_1,m_2-1,m_3;3}^z + s_{m_1,m_2,m_3-1;4}^z\right).
\label{a01}
\end{align}
Within the Tyablikov approximation $s^zs^{\pm}\to \langle s^z\rangle s^{\pm}$ 
(obviously, 
two spin operators here are attached to different sites
and
$\langle s^z\rangle$ is already site independent)
and, therefore,
\begin{eqnarray}
\label{a02}
\frac{1}{\sqrt{{\cal{N}}}}\sum_m{\re}^{-{\rm{i}}{\bf{q}}\cdot{\bf{R}}_m}\left[s_{m;1}^+,H\right]
\to
-2J\langle s^z\rangle
\left(3s^+_{{\bf{q}};1} -\phi_1 s^+_{{\bf{q}};2} -\phi_2 s^+_{{\bf{q}};3} -\phi_3 s^+_{{\bf{q}};4}\right);
\end{eqnarray}
for the definition of $\phi_a$ see equation~(\ref{206}).
As a result, according to equation~(\ref{303}), we obtain
\vspace{-5mm}
\begin{eqnarray}
\label{a03}
\omega G_{1\beta}(\omega)
=
2\langle s^z\rangle\delta_{1\beta}
-2J \langle s^z\rangle
\left[3 G_{1\beta}(\omega) -\phi_1 G_{2\beta}(\omega) -\phi_2 G_{3\beta}(\omega) -\phi_3 G_{4\beta}(\omega)\right].
\end{eqnarray}
Setting $\alpha=2$, $\alpha=3$, and $\alpha=4$ and repeating such calculations thrice, we obtain three more equations.
After all,
we may combine them in the matrix form 
\begin{eqnarray}
\label{a04}
\left(\omega{\bf{1}}+2J\langle s^z\rangle {\bf{F}}\right){\bf{G}}(\omega)=2\langle s^z\rangle {\bf{1}},
\end{eqnarray}
where the matrix ${\bf{F}}$ is defined in equation~(\ref{206}),
cf. equation~(\ref{304}).

It is worth commenting briefly on the mean-field approximation.
Within the frames of this approximation we replace equation~(\ref{201}) 
by $H=-3NJ\langle s^z\rangle^2 +6J\langle s^z\rangle\sum_m(s_{m;1}^z+s_{m;2}^z+s_{m;3}^z+s_{m;4}^z)$.
After that,
the exact first-order equation (\ref{303})
$\omega G_{\alpha\beta}(\omega)=2\langle s^z\rangle\delta_{\alpha\beta}-6J\langle s^z\rangle G_{\alpha\beta}(\omega)$
immediately yields
$G_{\alpha\beta}(\omega)=2\langle s^z\rangle\delta_{\alpha\beta}/\left(\omega+6J\langle s^z\rangle\right)$
that results in
$\langle s^-_{{\bf{q}};\,\beta} s^+_{{\bf{q}};\alpha}\rangle
=2\langle s^z\rangle\delta_{\alpha\beta}/({\re}^{6\vert J\vert\langle s^z\rangle/T}-1)$,
cf. equations~(\ref{305}) and (\ref{306}).
The self-consistent equation for $\langle s^z\rangle$ reads:
$1=2 \langle s^z\rangle\coth(3\vert J\vert \langle s^z\rangle/T)$
and it yields $T_{\text c}=\frac{3}{2}\vert J\vert$,
cf. equations~(\ref{307}) and (\ref{401}).
For the internal energy, we have:
$\frac{\langle H\rangle}{N}=-3\vert J\vert\langle s^z\rangle^2$,
cf. equation~(\ref{308}).
The inverse susceptibility above $T_{\text c}$ is as follows:
$1/\chi=4(T_{\text c}-T)$,
cf. equation~(\ref{311}).
Clearly, the mean-field results for observables are identical for the pyrochlore lattice and for the simple-cubic lattice.

Let us explain how to calculate the internal energy within the Tyablikov approximation.
The internal energy is defined as the average of Hamiltonian (\ref{201}),
i.e., as $\langle H\rangle$.
To account the transverse correlation energy and the longitudinal correlation energy on equal footing,
let us obtain first a formula for $\langle H\rangle$ 
that will contain only such correlation functions which are evaluated straightforwardly using the same Tyablikov approximation.
We begin with calculating the commutators $[s^+_{n;\alpha},H]$, $\alpha=1,2,3,4$, see equation~(\ref{a01}),
which yield the correlation functions $\langle s_{n;\alpha}^-[s^+_{n;\alpha},H]\rangle$, $\alpha=1,2,3,4$.
We use the relations $s^-s^z= s^- + s^zs^-$ and $s^-s^+=\frac{1}{2}-s^z$ 
to arrive at the wanted identity with $\langle H\rangle$:
\begin{align}
&\sum_{\alpha=1}^4\sum_{n=1}^{\cal{N}} \langle s_{n;\alpha}^-[s^+_{n;\alpha},H]\rangle
=
-12{\cal{N}}J\langle s^z\rangle +2\langle H\rangle
\nonumber\\
&+J\sum_n
\Big[
\big\langle s_{n;1}^zs_{n;1}^- 
\big(s^+_{n;2} + s^+_{n;3} + s^+_{n;4} + s^+_{n_1-1;2} + s^+_{n_2-1;3} + s^+_{n_3-1;4}\big)\big\rangle
\nonumber\\
&+
\big\langle s_{n;2}^zs_{n;2}^- 
\big(s^+_{n;1} + s^+_{n;3} + s^+_{n;4} + s^+_{n_1+1;1} + s^+_{n_1+1,n_2-1;3} + s^+_{n_1+1,n_3-1;4}\big)\big\rangle
\nonumber\\
&+
\big\langle s_{n;3}^zs_{n;3}^- 
\big(s^+_{n;1} + s^+_{n;2} + s^+_{n;4} + s^+_{n_2+1;1} + s^+_{n_1-1,n_2+1;2} + s^+_{n_2+1,n_3-1;4}\big)\big\rangle
\nonumber\\
&+
\big\langle s_{n;4}^zs_{n;4}^- 
\big(s^+_{n;1} + s^+_{n;2} + s^+_{n;3} + s^+_{n_3+1;1} + s^+_{n_1-1,n_3+1;2} + s^+_{n_2-1,n_3+1;3}\big)\big\rangle
\Big]
\label{a05}
\end{align}
(to keep notations short, we have suppressed some site indexes)
or using the ${\bf{q}}$-momentum representation, see equation~(\ref{301}),
\begin{align}
&\sum_{\alpha=1}^4 \sum_{\bf{q}} \langle s_{{\bf{q}};\alpha}^-[s^+_{{\bf{q}};\alpha},H]\rangle
=
-12{\cal{N}}J\langle s^z\rangle +2\langle H\rangle
\nonumber\\
&+J\sum_{\bf{q}}
\left[
\big\langle (s^zs^-)_{{{\bf{q}};1}}  s^+_{{\bf{q}};2}\big\rangle \big(1+{\re}^{-{\rm{i}}q_1}\big)
+
\big\langle (s^zs^-)_{{{\bf{q}};1}}  s^+_{{\bf{q}};3}\big\rangle \big(1+{\re}^{-{\rm{i}}q_2}\big)
+
\big\langle (s^zs^-)_{{{\bf{q}};1}}  s^+_{{\bf{q}};4}\big\rangle \big(1+{\re}^{-{\rm{i}}q_3}\big)
\right.
\nonumber\\
&+
\big\langle (s^zs^-)_{{{\bf{q}};2}}  s^+_{{\bf{q}};1}\big\rangle \big(1+{\re}^{{\rm{i}}q_1}\big)
+
\big\langle (s^zs^-)_{{{\bf{q}};2}}  s^+_{{\bf{q}};3}\big\rangle \big(1+{\re}^{{\rm{i}}q_1-{\rm{i}}q_2}\big)
+
\big\langle (s^zs^-)_{{{\bf{q}};2}}  s^+_{{\bf{q}};4}\big\rangle \big(1+{\re}^{{\rm{i}}q_1-{\rm{i}}q_3}\big)
\nonumber\\
&+
\big\langle (s^zs^-)_{{{\bf{q}};3}}  s^+_{{\bf{q}};1}\big\rangle \big(1+{\re}^{{\rm{i}}q_2}\big)
+
\big\langle (s^zs^-)_{{{\bf{q}};3}}  s^+_{{\bf{q}};2}\big\rangle \big(1+{\re}^{-{\rm{i}}q_1+{\rm{i}}q_2}\big)
+
\big\langle (s^zs^-)_{{{\bf{q}};3}}  s^+_{{\bf{q}};4}\big\rangle \big(1+{\re}^{{\rm{i}}q_2-{\rm{i}}q_3}\big)
\nonumber\\
&\left.
+
\big\langle (s^zs^-)_{{{\bf{q}};4}}  s^+_{{\bf{q}};1}\big\rangle \big(1+{\re}^{{\rm{i}}q_3}\big)
+
\big\langle (s^zs^-)_{{{\bf{q}};4}}  s^+_{{\bf{q}};2}\big\rangle \big(1+{\re}^{-{\rm{i}}q_1+{\rm{i}}q_3}\big)
+
\big\langle (s^zs^-)_{{{\bf{q}};4}}  s^+_{{\bf{q}};3}\big\rangle \big(1+{\re}^{-{\rm{i}}q_2+{\rm{i}}q_3}\big)
\right]
\nonumber\\
&=
-12{\cal{N}}J\langle s^z\rangle +2\langle H\rangle
+2J\sum_{\bf{q}}
\left[
\big\langle (s^zs^-)_{{{\bf{q}};1}}  s^+_{{\bf{q}};2}\big\rangle \phi_1
+
\big\langle (s^zs^-)_{{{\bf{q}};1}}  s^+_{{\bf{q}};3}\big\rangle \phi_2
+
\big\langle (s^zs^-)_{{{\bf{q}};1}}  s^+_{{\bf{q}};4}\big\rangle \phi_3
\right.
\nonumber
\\
&+
\big\langle (s^zs^-)_{{{\bf{q}};2}}  s^+_{{\bf{q}};1}\big\rangle \phi_1^*
+
\big\langle (s^zs^-)_{{{\bf{q}};2}}  s^+_{{\bf{q}};3}\big\rangle \phi_{12}^*
+
\big\langle (s^zs^-)_{{{\bf{q}};2}}  s^+_{{\bf{q}};4}\big\rangle \phi_{13}^*
\nonumber
\\
%\end{align}
%\begin{align}
&+
\big\langle (s^zs^-)_{{{\bf{q}};3}}  s^+_{{\bf{q}};1}\big\rangle \phi_2^*
+
\big\langle (s^zs^-)_{{{\bf{q}};3}}  s^+_{{\bf{q}};2}\big\rangle \phi_{12}
+
\big\langle (s^zs^-)_{{{\bf{q}};3}}  s^+_{{\bf{q}};4}\big\rangle \phi_{23}^*
\nonumber
\\
&\left.
+
\big\langle (s^zs^-)_{{{\bf{q}};4}}  s^+_{{\bf{q}};1}\big\rangle \phi_3^*
+
\big\langle (s^zs^-)_{{{\bf{q}};4}}  s^+_{{\bf{q}};2}\big\rangle \phi_{13}
+
\big\langle (s^zs^-)_{{{\bf{q}};4}}  s^+_{{\bf{q}};3}\big\rangle \phi_{23}
\right];
\label{a06}
\end{align}
for definition of $\phi_a$ and $\phi_{ab}$ see equation~(\ref{206}).
In the left-hand side of equation~\eqref{a06} we can obviously write
\begin{eqnarray}
\label{a07}
\langle s_{{\bf{q}};\alpha}^-[s^+_{{\bf{q}};\alpha},H]\rangle
=
{\rm{i}}\left[\frac{{\rm{d}}}{{\rm{d}} t} \big\langle s_{{\bf{q}};\alpha}^- s^+_{{\bf{q}};\alpha}(t)\big\rangle \right]_{t=0}
=
{\rm{i}}\left[\frac{{\rm{d}}}{{\rm{d}} t} 
\frac{1}{2\piup}\int_{-\infty}^\infty{\rm{d}}\omega {\re}^{-{\rm{i}}\omega t} J_{\alpha\alpha}(\omega) \right]_{t=0}
=
\frac{1}{2\piup}\int_{-\infty}^\infty{\rm{d}}\omega \omega J_{\alpha\alpha}(\omega),
\end{eqnarray}
where
\begin{eqnarray}
\label{a08}
J_{\alpha\alpha}(\omega)
=
{\rm{i}}\frac{G_{\alpha\alpha}(\omega+{\rm{i}}0)-G_{\alpha\alpha}(\omega-{\rm{i}}0)}{{\re}^{\frac{\omega}{T}}-1}
=
2\piup
\sum_\gamma
\frac{2\langle s^z\rangle\langle\alpha\vert\gamma{\bf{q}}\rangle\langle\gamma{\bf{q}}\vert\alpha\rangle}
{{\re}^{\frac{\omega}{T}}-1}\delta\left(\omega+2J\langle s^z\rangle\omega_{\gamma;{\bf{q}}}\right);
\end{eqnarray}
we have used equation~(\ref{305}).
That is,
\begin{eqnarray}
\label{a09}
\sum_\alpha
\sum_{\bf{q}} \langle s_{{\bf{q}};\alpha}^-[s^+_{{\bf{q}};\alpha},H]\rangle
=
4\vert J\vert\langle s^z\rangle^2\sum_{\gamma=1}^4 \sum_{\bf{q}}
\frac{\omega_{\gamma;{\bf{q}}}}{{\re}^{\frac{2\vert J\vert\langle s^z\rangle\omega_{\gamma;{\bf{q}}}}{T}}-1}.
\end{eqnarray}
In the right-hand side of equation~\eqref{a06} we have the correlation functions 
$\langle (s^zs^-)_{{{\bf{q}};\,\beta}}  s^+_{{\bf{q}};\alpha}\rangle$
which are calculated from the Green's function 
$H_{\alpha\beta}(t)\equiv\langle\langle s^+_{{\bf{q}};\alpha}\vert (s^zs^-)_{{{\bf{q}};\,\beta}}\rangle\rangle_t
=-{\rm{i}}\Theta(t)\langle [s^+_{{\bf{q}};\alpha}(t), (s^zs^-)_{{{\bf{q}};\,\beta}}]\rangle $,
$H_{\alpha\beta}(\omega)=\int_{-\infty}^\infty{\rm{d}}t {\re}^{{\rm{i}}\omega t}H_{\alpha\beta}(t)$
using the spectral representation
\begin{eqnarray}
\label{a10}
\langle (s^zs^-)_{{{\bf{q}};\,\beta}}  s^+_{{\bf{q}};\alpha}\rangle
=
\frac{{\rm{i}}}{2\piup}
\int_{-\infty}^\infty{\rm{d}}\omega
\frac{H_{\alpha\beta}(\omega+{\rm{i}}0)- H_{\alpha\beta}(\omega-{\rm{i}}0)}{{\re}^{\frac{\omega}{T}}-1}\,,
\end{eqnarray}
cf. equation~(\ref{306}).
Repeating the derivation of equation~(\ref{a04}) 
(equation of motion within the Tyablikov approximation),
but now for the Green's function ${\bf{H}}(\omega)$,
we arrive at
\begin{eqnarray}
\label{a11}
\sum_\gamma\left(\omega\delta_{\alpha\gamma}
+
2J\langle s^z\rangle F_{\alpha\gamma}\right)H_{\gamma\beta}(\omega)
=
-\langle s^z\rangle\delta_{\alpha\beta}\,,
\end{eqnarray}
cf. equations~\eqref{304} and (\ref{a04}), 
and as a result we obtain
\begin{eqnarray}
\label{a12}
H_{\alpha\beta}(\omega)
=
-\langle s^z\rangle
\sum_{\gamma=1}^4\frac{\langle\alpha\vert \gamma{\bf{q}}\rangle\langle \gamma{\bf{q}}\vert\beta\rangle}
{\omega+2J\langle s^z\rangle\omega_{\gamma;{\bf{q}}}}\,,
\end{eqnarray}
cf. equation~(\ref{305}). 
Therefore,
\begin{eqnarray}
\label{a13}
\langle (s^zs^-)_{{{\bf{q}};\,\beta}}  s^+_{{\bf{q}};\alpha}\rangle
=
- \langle s^z\rangle 
\sum_{\gamma=1}^4
\frac{\langle\alpha\vert \gamma{\bf{q}}\rangle\langle \gamma{\bf{q}}\vert\beta\rangle}
{{\re}^{\frac{2\vert J\vert\langle s^z\rangle\omega_{\gamma;{\bf{q}}}}{T}}-1}\,,
\end{eqnarray}
cf. equation~(\ref{306}).
Inserting equation~(\ref{a09}) and equation~(\ref{a13}) into equation~(\ref{a06}),
we obtain
\begin{align}
&4\vert J\vert\langle s^z\rangle^2\!\sum_{\gamma=1}^4 \!\sum_{\bf{q}}
\!\frac{\omega_{\gamma;{\bf{q}}}}{{\re}^{\frac{2\vert J\vert\langle s^z\rangle\omega_{\gamma;{\bf{q}}}}{T}}-1}
\!=\!
3N\vert J\vert\langle s^z\rangle \!+\!2\langle H\rangle
\!+\!2\vert J\vert  \langle s^z\rangle  
\!\sum_{\gamma=1}^4
\!\sum_{\substack{\alpha,\beta=1\\(\alpha\ne\beta)}}^4
\!\sum_{\bf{q}}\varphi_{\alpha\beta}
\frac{\langle\alpha\vert \gamma{\bf{q}}\rangle\langle \gamma{\bf{q}}\vert\beta\rangle}
{{\re}^{\frac{2\vert J\vert\langle s^z\rangle\omega_{\gamma;{\bf{q}}}}{T}}-1}\,,
\nonumber\\
&\varphi_{12}=\varphi_{21}^*=\phi_1\,, \qquad
\varphi_{13}=\varphi_{31}^*=\phi_2\,,\qquad
\varphi_{14}=\varphi_{41}^*=\phi_3\,,
\nonumber\\
&\varphi_{23}=\varphi_{32}^*=\phi_{12}^*\,, \qquad
\varphi_{24}=\varphi_{42}^*=\phi_{13}^*\,, \qquad
\varphi_{34}=\varphi_{43}^*=\phi_{23}^*.
\label{a14}
\end{align}
This is equivalent to equation~\eqref{308}.
Note also that the expression for the internal energy (\ref{308}) agrees with the formula given in equation~(336) of \cite{Froebrich2006}.

\ukrainianpart

\title{Спін-$\frac{1}{2}$ феромагнетик Гайзенберга на гратці пірохлору. Дослідження методом функцій Гріна}
\author{Т. Гутак\refaddr{label1},
        П. Мюллер\refaddr{label2}, 
        Й. Ріхтер\refaddr{label2,label3},
        Т. Крохмальський\refaddr{label1},
        О. Держко\refaddr{label1,label3}}
\addresses{
\addr{label1} Інститут фізики конденсованих систем НАН України, 
        вул. Свєнціцького, 1, 79011 Львів, Україна
\addr{label2} Інститут теоретичної фізики, Магдебурзький університет імені Отто фон Ґеріке, \\
        поштова скринька 4120, 39016 Магдебург, Німеччина
\addr{label3} Інститут Макса Планка фізики складних систем, 
        Ньотнітцерштрасе 38, 01187 Дрезден, Німеччина
}

\makeukrtitle

\begin{abstract}
\tolerance=3000%
Ми розглядаємо квантовий феромагнетик Гайзенберга на гратці пірохлору 
і обговорюємо властивості цієї спінової моделі при довільних температурах.
Для цього ми використовуємо техніку функцій Гріна у наближенні хаотичних фаз (наближенні Тяблікова), 
а також лінійну теорію спінових хвиль і симуляції методом квантового Монте Карло.
Для підтвердження наших результатів ми порівнюємо їх з  отриманими недавно іншими способами.
Нарешті ми співставляємо наші результати з відповідними для простої кубічної гратки, 
адже обидві гратки ідентичні у середньопольовій картині.
Ми показуємо, що теплові флуктуації ефективніші у випадку гратки пірохлору
(ефекти фрустрації при скінченних температурах).
Наші результати можна використати для тлумачення експериментальних даних для матеріалів, у яких реалізується феромагнетик на гратці пірохлору.

\keywords квантова спінова система, модель Гайзенберга, гратка пірохлору, функції Гріна, наближення хаотичних фаз

\end{abstract}

\end{document}